\input{epsfig.sty}
\documentstyle[aps,12pt,manuscript]{revtex}

\def\makefront{
\hfill OCHA-PP-122 \\
\vspace*{1cm}\begin{center}
\def\newtitleline{\\ \vskip 5pt}
{\Large\bf\titleline}\\
\vskip 1truecm
{\large\bf\authors}\\
\vskip 5truemm
\addresses
\end{center}
\vskip 1truecm
{\bf Abstract:}
\abstracttext
\vskip 1truecm}
\begin{document}
\def\titleline{
Genetic Algorithm for SU(N) gauge theory on a lattice
}

\def\authors{
  A. Yamaguchi \footnote{ E-mail address: azusa@sokrates.phys.ocha.ac.jp}  
}

\def\addresses{
The Particle Physics Laboratory \\
Ochanomizu University \\
2-1-1 Otsuka Bunkyoku Tokyo, Japan\\
 }

\def\abstracttext{
An algorithm is proposed for the simulation of pure SU(N) 
lattice gauge theories based on Genetic Algorithms(GAs). 
Main difference between GAs and  Metropolis methods(MPs) is that
GAs treat a population of points at once,
while MPs treat only one point in the searching space.
This provides GAs with information about the assortment as well as
the fitness of the evolution function 
and producing a better solution. We apply GAs to SU(2) pure gauge
theory on a 2 dimensional lattice and 
show the results are consistent with those by MPs
and Heatbath methods(HBs). Thermalization speed of GAs is especially 
faster than the simple MPs
}

\makefront

\section{Introduction}

Genetic Algorithm is based upon the theory of evolution.
The study of GAs originates with John Holland~\cite{Holland} in the mid-1970s.
Since the mid-1980s GAs
have been explored how to use in a various fields of applied sciences
and industrial technology
for optimizations (ex. Traveling Salesman problem)
 stochastic search (ex. Pattern matching)
and learning system (ex. Neural Network).
 
GA is defined as a computer simulation in which
by a change of generations with selections and multiplications
depending on the fitness value of the evaluation of a population
of virtual organisms on computer,
better genetics of a previous generations tends to be
passed on its offspring. Here selection means the survival of the fittest,
while multiplications are the processes in that organisms multiply.
 
GAs work best in the following situations~\cite{EPCC}:
 
\begin{enumerate}
 
\item Potential solutions can be represented in a way which exposes components
of solutions, and
 
\item operators to mutate and hybridize these representations are available.
 
\end{enumerate}

GAs have two advantage points that updating process
need neither any local data nor global data and that
the fitness is only evaluated after updating process
is guided to a good solution in a searching space.
 
In lattice gauge field theory, an individual in GAs corresponds to
a whole lattice and each field variables 
corresponds to chromosome.
The evaluation function is the normal lattice action. 
 
Main procedures of GAs are selection, multiplication, mutation and crossover.
Among them, mutation is the same idea as updating process of Metropolis
method, but MPs use mutation alone.
HBs and MPs treat only one lattice configuration, whereas
GAs can treat a various number of lattice configurations. 

We applied GAs to pure SU(2) gauge theory in 2-dimensions, 
explored various schemes and examined its effectiveness in the simplest way.
Our results show the distinctive features of GAs which lead to the
 fast thermalization.
The short calculation time are accomplished by the encoding   
configurations on a lattice.

Even if individuals are systems including fermion fields which bring
the difficulty of the non-local logarithmic determinants,
GAs might be able to simulate without particular difficulty.

\section{
Genetic Algorithms for SU(N)}

The success of GAs strongly depends on how to map 
the concrete representations toward the abstract searching space in
which operators move through.

Since integer is faster than real operation ,
the information of field configurations are encoded to integer values of which
bit patterns are treated as chromosome.

After encoding procedures, the current genomes are recombined and produce new
genomes which might be able to produce next offsprings.
Among the schemes at the recombination stage, we have two kinds of 
mutation schemes with ensure ergodicity. 
If all the 1 bits in the particular position of genome
are eliminated, a 1 bit will eventually reappear there by the mutation 
for genome. Therefore the coverage of the whole searching space are ensured 
by the mutation for the source of information of configurations.

GAs' procedures, crossover
and mutation, tend to make candidate solutions to climb hills and
escape from the local minima, i.e. GAs are global search compared
with MPs.
After global search by GAs, the local search by Metropolis
methods become effective for identifying the true minima of solutions.
At this stage the solution are already converged around the best solutions and 
a local search do not break
down them. This method is called Hybrid GAs.
 
Our combination scheme of two kind of mutations means not only 
the satisfaction of ergodicity but also our algorithm is
Hybrid Genetic Algorithm.
It has been known that Hybrid GAs are good at Traveling Salesman Problem
~\cite{Malek}.

A flowchart of a Hybrid GA for SU(N) is given as

\begin{quote}
\baselineskip=10pt
Hybrid GAs  \\
{\bf INITIALIZATION} \\
 creating link variables \\
 coding phenotype to genotype (binary strings) \\
\\
{\bf REPRODUCTION } \\
 Selection and Multiplication \\
 Crossover and Mutation \\
 Evaluation \\
\\
{\bf METROPOLIS METHOD}
\\
{\bf the ultima generation} \\
\end{quote}

At the reproduction stage, if there are no clear difference between fitness 
values of individuals, the difference is emphasized by Scaling, or
if there are too clear difference between fitness values, Scaling make them 
obscure.  Basically, there are three kinds of Scaling functions, Linear scaling,
Exponential scaling and Sigma scaling (Windowing).
 
\begin{center}
\begin{tabular}{ll}
  Linear Scaling & $ f_i ' = af_i + b $ \\
  Exponential Scaling & $ f_i ' = (f_i)^k $ \\
  Sigma Scaling ( Windowing) & $ f_i ' = f_i - (\bar{f}-c\times \sigma)$ \\
\end{tabular}
\end{center}
where $f_i$ is the raw fitness value of i'th genome, $\bar{f}$ is the mean,
and $\sigma$ is the standard deviation with some integer k and 
some constants a, b and c. 
 
In Sigma Scaling, genomes with fitness value lower than $c \times \sigma$
are treated as fatal ones.

The selection is based on the survival of the fittest.
Genomes of the previous generation are selected as candidates for parents,
according to their own scaled/raw fitness values.
It is important, however that even genomes with lower values have
some possibility to pass their genetics on the next generation, because
it might help the system to escape from local minima.
Generally there are three kinds of schemes of selection
expected to select genomes perfect as possible, Roulette Selection,
Tournament Selection, and Stochastic Universal Sampling.
Among them, Stochastic Universal Sampling which chooses
N genomes from the population at once
ensures a sampling including low fitness value genomes.
Since a genome on the top of the stack by this sampling takes
to be a candidate of parents, this scheme is the most efficient way to
maintain diversity of children.

Reproduction schemes are Crossover and Mutation. In SU(2) pure gauge
theory, Crossover means to exchange the elements of gauge field variables
on each lattice point. This scheme brings its action to increase, caused by
the increase of surface energy. The amount of increase is larger than
the expected decrease coming from the lowering of interior energy.
 
We use two schemes of Crossover. One is Uniform Crossover in which a 
child genome is created with its own genetics that consists of genetics
passed from one of parents with some probability $p$ and those from
another one with the probability $1-p$. Each genotype is occupied by
the genes from parents independently.
Another scheme is 2-point Crossover that parents genomes whose forms are not 
strings but rings are split into 2 parts
at two crosspoints chosen at random, and then combined to make a new genome.

The different point between them is that 2-point Crossover might suppress some
increase of its action rather than Uniform Crossover. Diversity is to have
high probability of a production of an interesting child genome
is, however, assured by Uniform Crossover rather than 2-point Crossover.

The problem which remains and to be expected is how to update the population.
Generally, N children produced from N parents replace all the parents, Then, the
next generation becomes totally new. This scheme is called spawning.
The diversity of the population depends, however, strongly on the selection
of the parents.
As Updating schemes Breeding and spawning are presented following  nature species.
Spawning is also called as Discrete generation scheme.
Swap all of the previous generation with the new generation.
Breeding is called as Continuous generation scheme in which parents and children
are mixed by the following three replacements; 
the replacement of parent that the child was produced from,
the replacement of the worst genome in the population,
or the replacement of the oldest genome in the population.
De Jong introduced a parameter called Generation gap which is the ratio of
the number of children to the number of parents in a generation~\cite{DeJong}.
The scheme called Elitsm might be thought as that it could improve
 Updating procedure,
because replacement of the best individual in a population is forbidden.
It must be remarked, however, that 
it has greater risk to fall into premature convergence than simple spawning and
breeding schemes.
In the same way, though schemes that replace 
the oldest/worst one or never replace the best one
also might be looked clever way, 
it yields no improvement to avoid false convergence at all.
 
The real problem we have is what our goal is. GAs search the space we create,
looking for the best solution. In our case, the final goal is a configuration
which stay in the thermal equilibrium, so that it is impossible
to establish thermal equilibrium
state without the detailed balance. It means that our updating scheme have
to ensure this constraint.
At the updating procedure, we set the accept/reject function between
a better parent and a better child. The transition probability
$P(\{\phi\}_{child} \leftarrow \{\phi\}_{parent})$  is generated  as
$ P=P_A P_C$ where
$P_C(\{\phi\}_{child} \leftarrow \{\phi\}_{parent})$ is an arbitrary probability
distribution for the proposed change from configurations of a parent to that of
a child, and $P_A$ is the acceptance probability 
$P_A(\{\phi\}_{child} \leftarrow \{\phi\}_{parent})$ are given by
\begin{equation}
  P_A(\{\phi\}_{child} \leftarrow \{\phi\}_{parent}) \propto
  \min \displaystyle{
   \left\{ 1, 
 \frac{ P_C(\{\phi\}_{parent} \leftarrow \{\phi\}_{child}) \cdot e^{S_{child}} }
  { P_C(\{\phi\}_{child} \leftarrow \{\phi\}_{parent})\cdot e^{S_{parent}} }
   \right\}. 
   }
\end{equation}
Here $S$ is the usual action of SU(N) lattice gauge theories, that is 
\begin{equation}
\displaystyle{
  S[U] = \beta\Sigma_p \left( 1 - \frac{1}{N}ReTrU_p \right),
             }
\end{equation}
where $U_p$ is the element of SU(N) defined on a plaquette $p$ and $\beta$ is
a coupling constant.

As the usual Metropolis algorithm, $P_{C}$ is corresponding to be uniform,
so that there is no bias to create particular configuration.

\section{Experimental results}

\begin{figure}[p]
\begin{center}
\epsfig{file=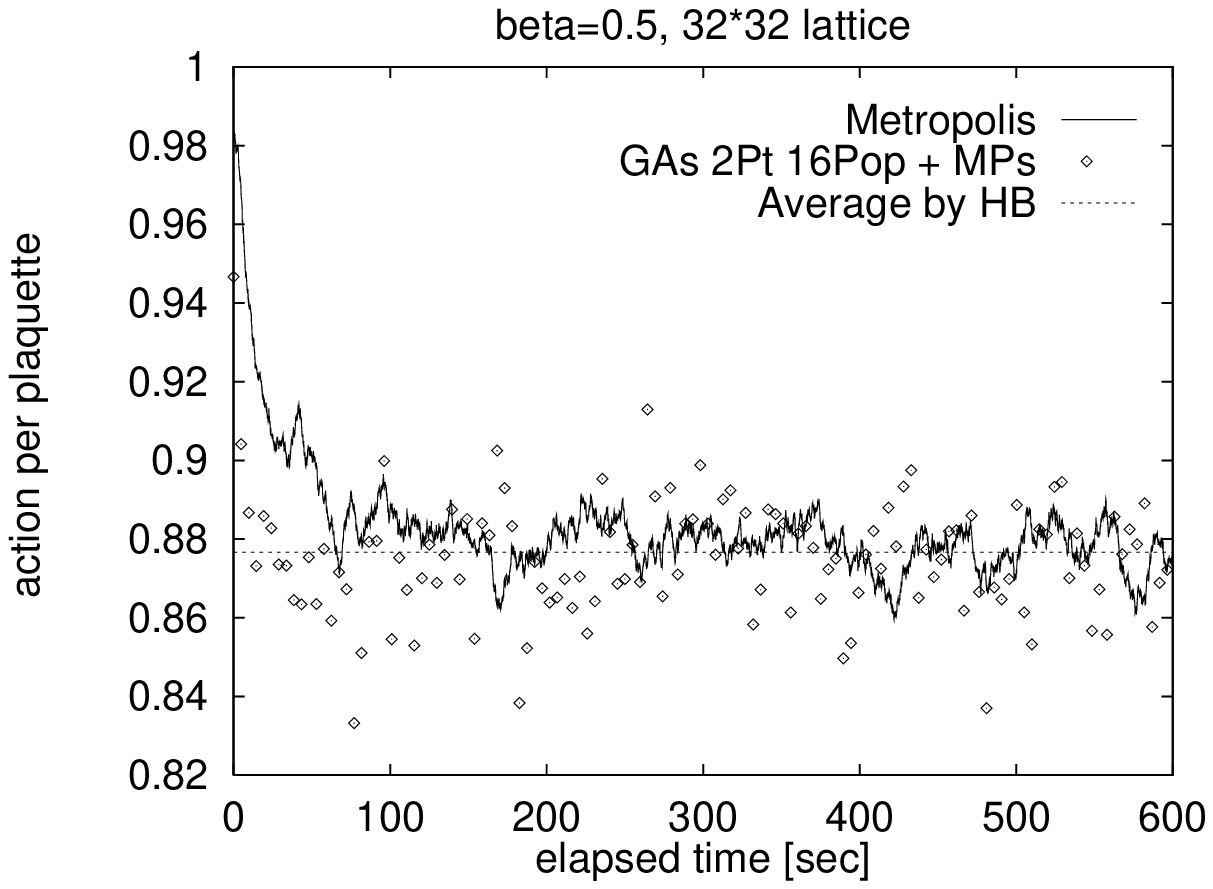,
        width=12.0cm,height=12.0cm,
        bbllx=50pt,bblly=50pt,bburx=350pt,bbury=350pt,
        angle=0}
\end{center}
\begin{center}
\parbox{12.5cm}{ \caption{ \label{fig1}
 Thermalization of the action per plaquette on $32 \cdot 32$ lattice
at$\beta=0.5$.}}
\end{center}
\end{figure}

\begin{figure}
\begin{center}
\epsfig{file=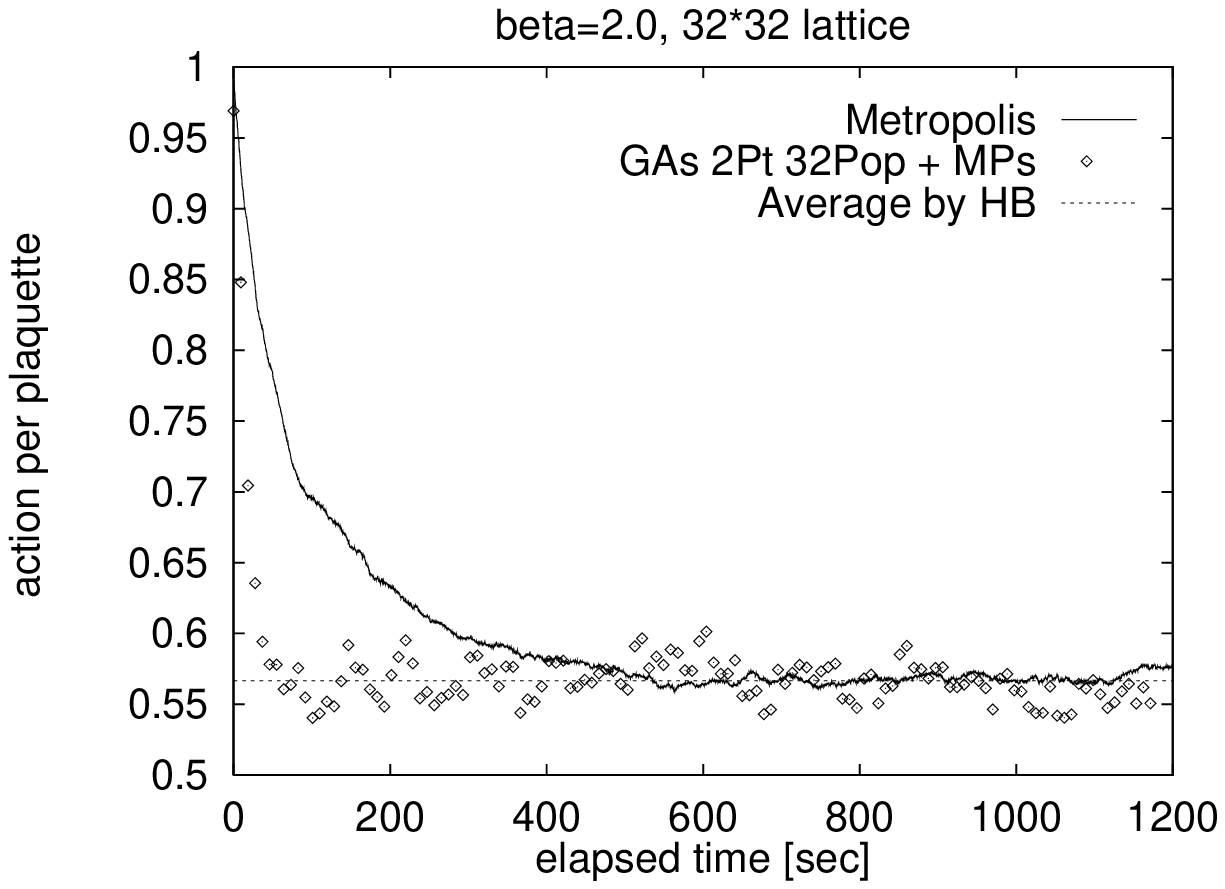,
        width=12cm,height=12.0cm,
        bbllx=50pt,bblly=50pt,bburx=350pt,bbury=350pt,
        angle=0}
\end{center}
\begin{center}
\parbox{12.5cm}{\caption{ \label{fig2}
 Thermalization of the action per plaquette on $32 \cdot 32$ lattice
at$\beta=2.0$.}}
\end{center}
\end{figure}

\begin{figure}
\begin{center}
\epsfig{file=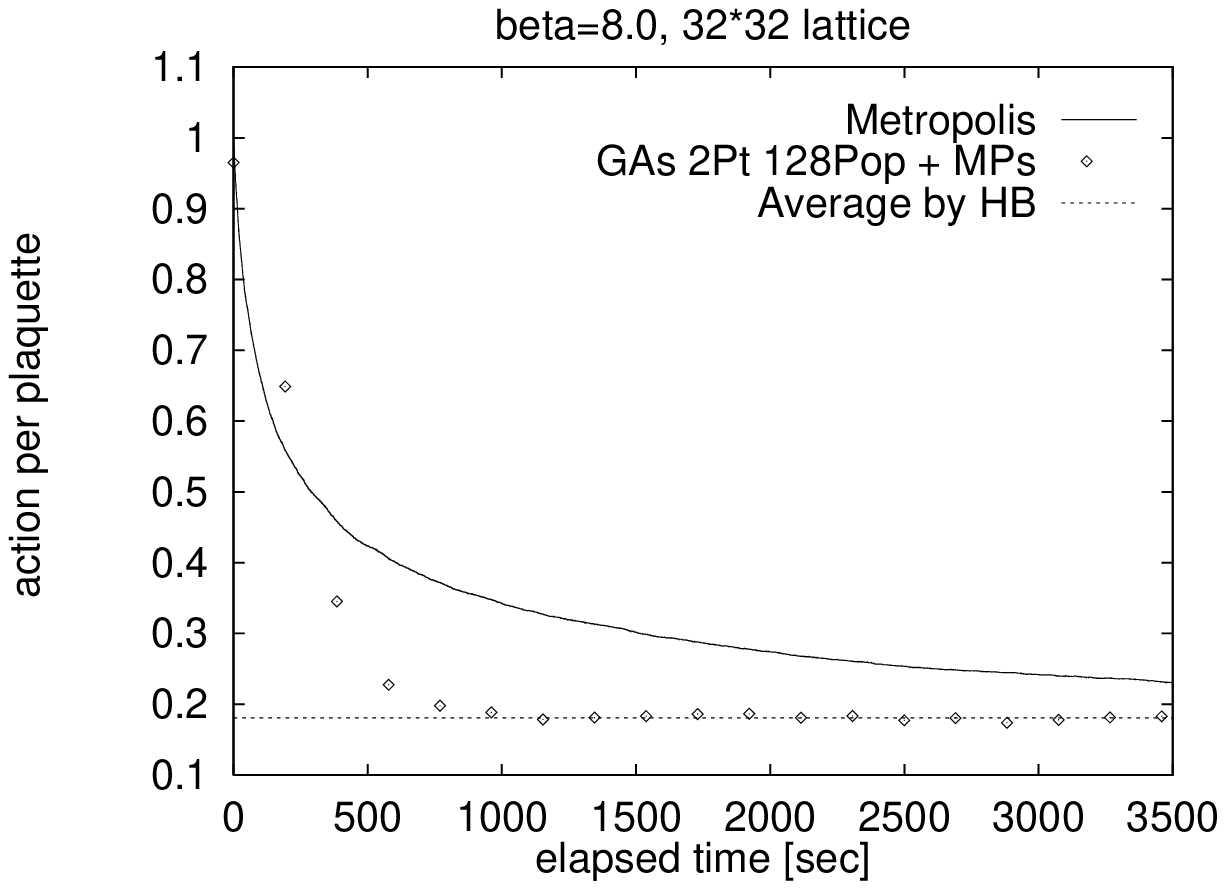,
        width=12cm,height=12.0cm,
        bbllx=50pt,bblly=50pt,bburx=350pt,bbury=350pt,
        angle=0}
\end{center}
\begin{center}
\parbox{12.5cm}{\caption{ \label{fig3}
 Thermalization of the action per plaquette on $32 \cdot 32$ lattice
at$\beta=8.0$.}}
\end{center}
\end{figure}

In this section, we present the simulation results and compare the thermalization of GAs with
that of MPs. 
In our GAs, Linear scaling and Stochastic universal sampling selection scheme
are used ,2point crossover scheme is adopted with crossrate $0.65$,
and Mutation ratio is
$0.008$. The population sizes are $16$ for  $\beta=0.5$, 
$32$ for $\beta =2.0$  and $128$ for $\beta = 8.0$. 
Since small $\beta$ has the large tolerance to
increase the action, it does not need a large population size.
Runs for $\beta$ values $0.2$, $0.5$ end at $128$th
generation, and a run for $\beta$ value $8.0$ ends at $18$th generation.
Their thermalization times are compared with those given by MPs.
Figures~\ref{fig1},~\ref{fig2} and ~\ref{fig3} 
show the results of thermalization 
with $\beta$ values $0.5$, $2.0$ and $8.0$, respectively.
In each figure, the horizontal dotted line shows the average value of 
action per plaquette of
the last $1000$ sweeps after $30,000$ sweeps, given by Heatbath method.
Square dots show values of the action per plaquette 
every generation,obtained by GAs. 
Time interval between generations depends on the population size.
Line show action per plaquette, given by MPs.
Runs with three kinds of $\beta$ values converge at the target value, 
the average values by Heatbath, faster than Simple Metropolis method. 
A run with a large $\beta$ converges quite fast,
instead of the low acceptance ratio.
The fluctuation of 
the run with $\beta = 0.5$ is, however, quite hard.
A bad crossover procedure among a low diversity population 
brings an unexpected increase of the action
that is accepted by the tolerance of a small $\beta$, otherwise
a good crossover procedure brings a large decrease of the action. 
Eventually an average value of action minima with a small $\beta$ 
is close to that of Heatbath.
Note however that while MPs treat just one lattice, 
GAs treats a population size number of lattices at once.%
 
\section{Simulation schemes}

There is a various way/process of evolution. How to combine them is
very important, since some wrong/strong combination leads to the premature
convergence at the local minima.
This section presents simulation results by comparing 2point crossover scheme 
and uniform crossover scheme, and by comparing the difference of a population
size.
Simulations are performed with two $\beta$, $0.2$ and $0.8$.
In their runs the uniform crossover with crossrate $0.65$
and 2point crossover with the same crossrate are used.
Population size are $32$ and $16$ for$\beta = 0.2$,
and $64$ and $128$ for$\beta = 8.0$.

\begin{figure}
\begin{center}
\begin{tabular}{ll}
\begin{minipage}{7.5cm}
\begin{center}
\epsfig{file=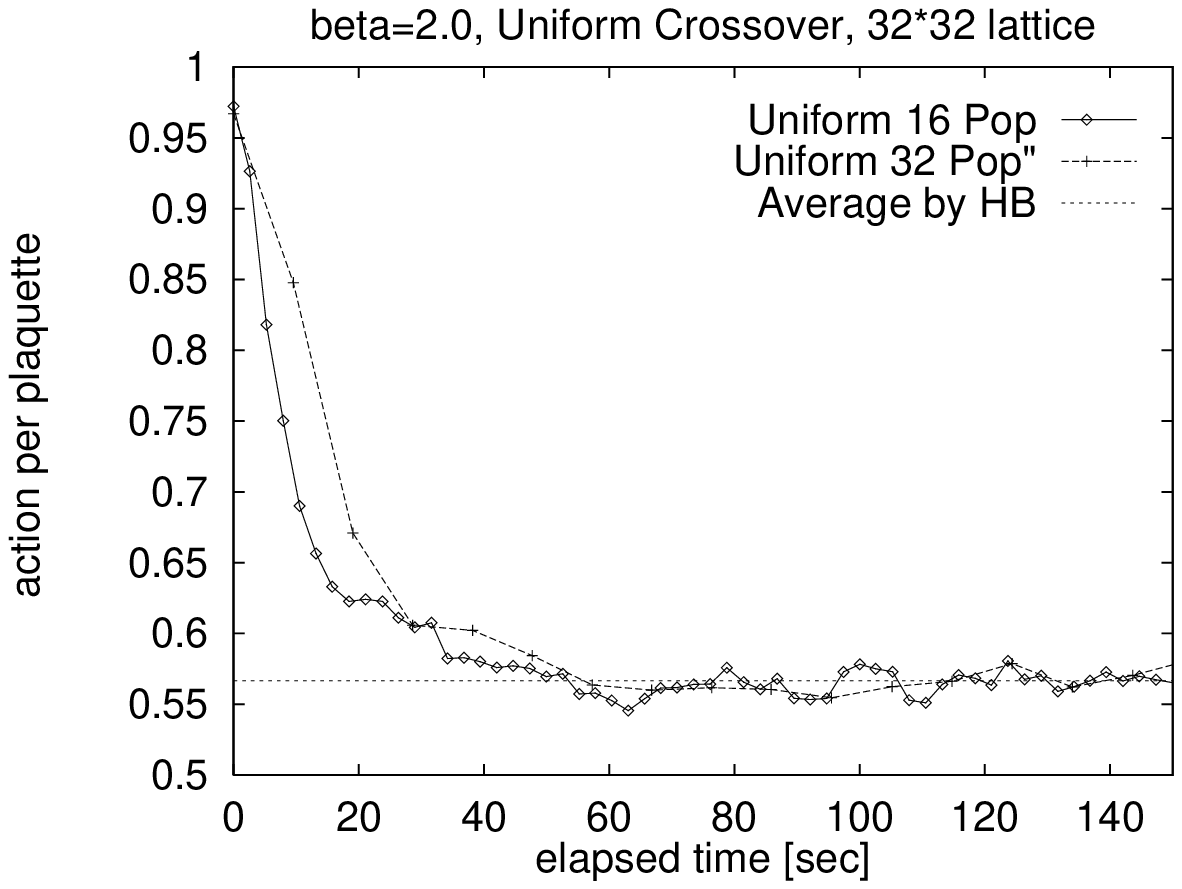,
        width=6.5cm,height=7.0cm,
        bbllx=40pt,bblly=40pt,bburx=340pt,bbury=340pt,
        angle=0}
\end{center}
\end{minipage}&
\begin{minipage}{7.5cm}
\begin{center}
\epsfig{file=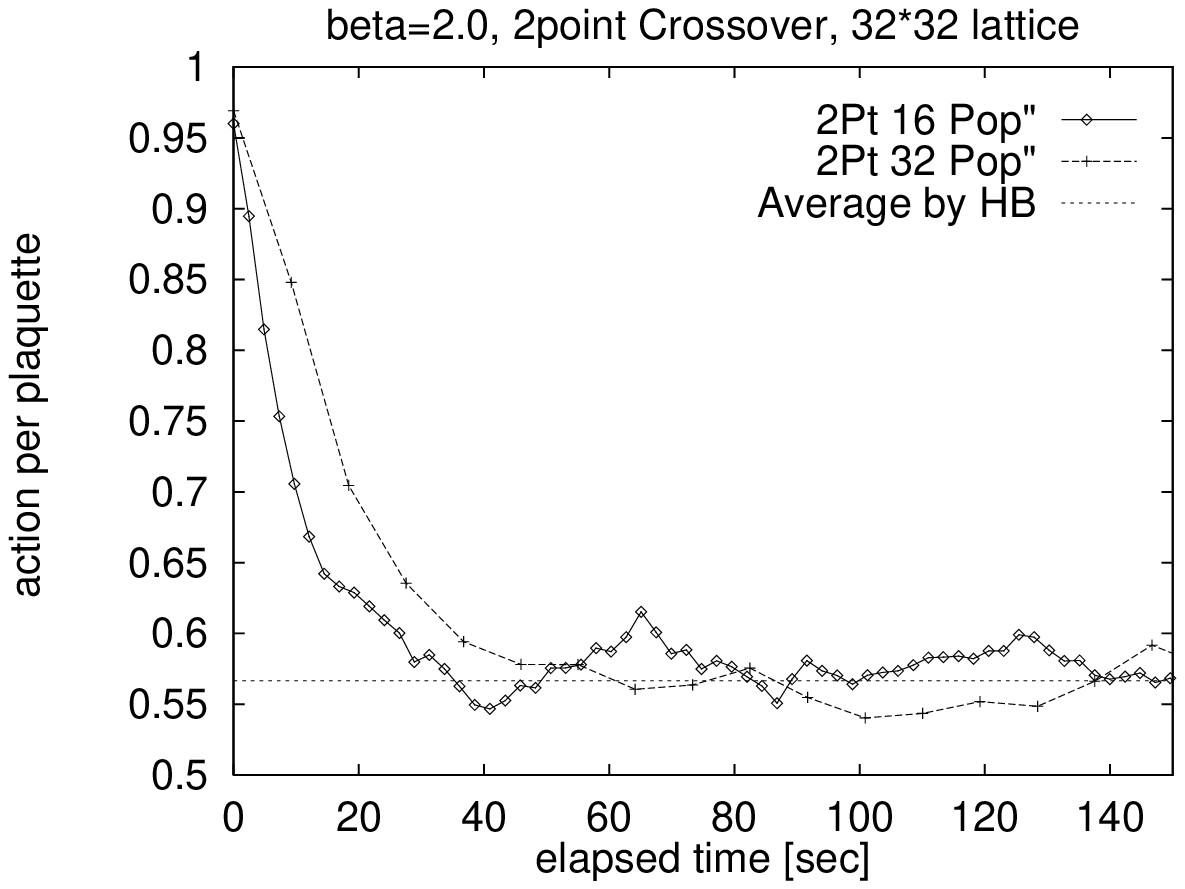,
        width=6.5cm,height=7.0cm,
        bbllx=40pt,bblly=40pt,bburx=340pt,bbury=340pt,
        angle=0}
\end{center}
\end{minipage} \\
\begin{minipage}{7.5cm}
\begin{center}
\epsfig{file=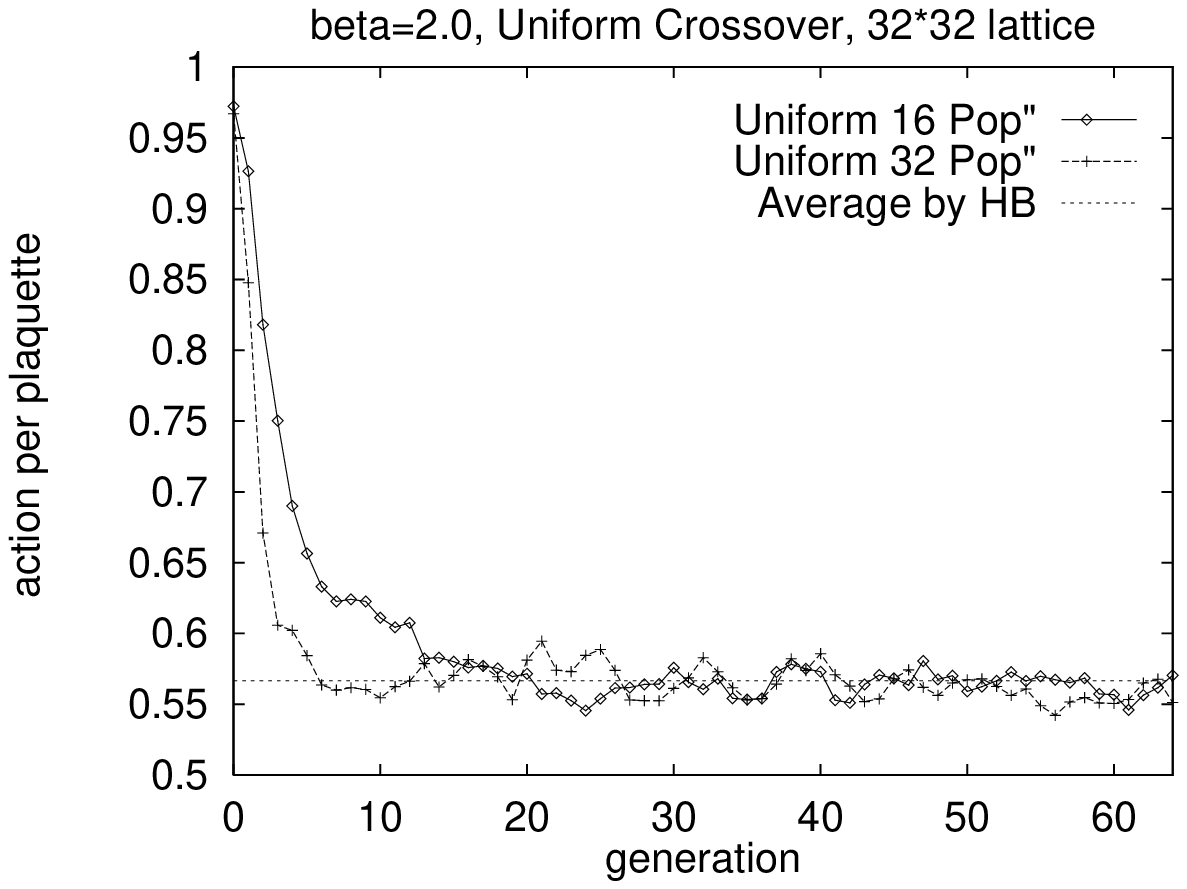,
        width=6.5cm,height=7.0cm,
        bbllx=40pt,bblly=40pt,bburx=340pt,bbury=340pt,
        angle=0}
\end{center}
\end{minipage}&
\begin{minipage}{7.5cm}
\begin{center}
\epsfig{file=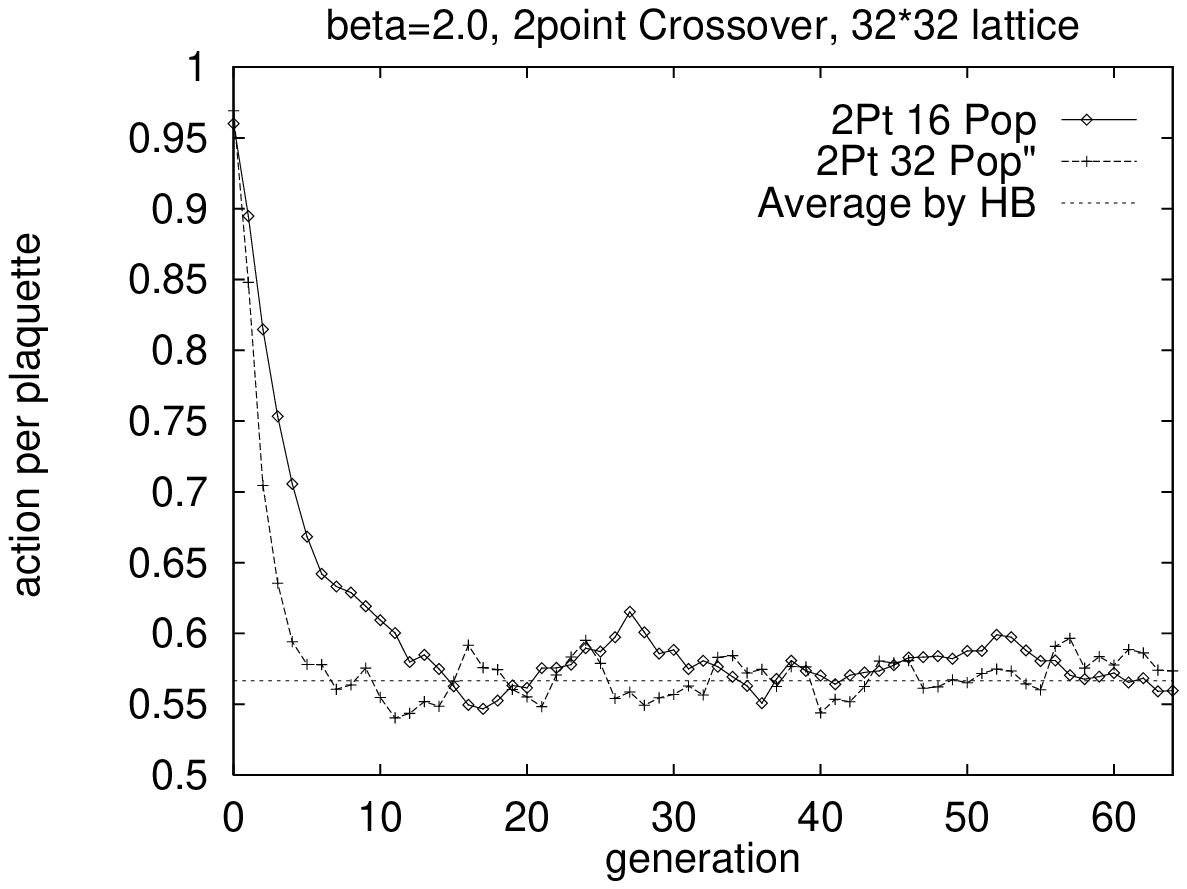,
        width=6.5cm,height=7.0cm,
        bbllx=40pt,bblly=40pt,bburx=340pt,bbury=340pt,
        angle=0}
\end{center}
\end{minipage}
\end{tabular}
\parbox{16cm}{ \caption{ \label{fig4}
Comparison of Recombination schemes and population sizes at $\beta=2.0$ }}
\end{center}
\end{figure}

\begin{figure}
\begin{center}
\begin{tabular}{ll}
\begin{minipage}{7.5cm}
\begin{center}
\epsfig{file=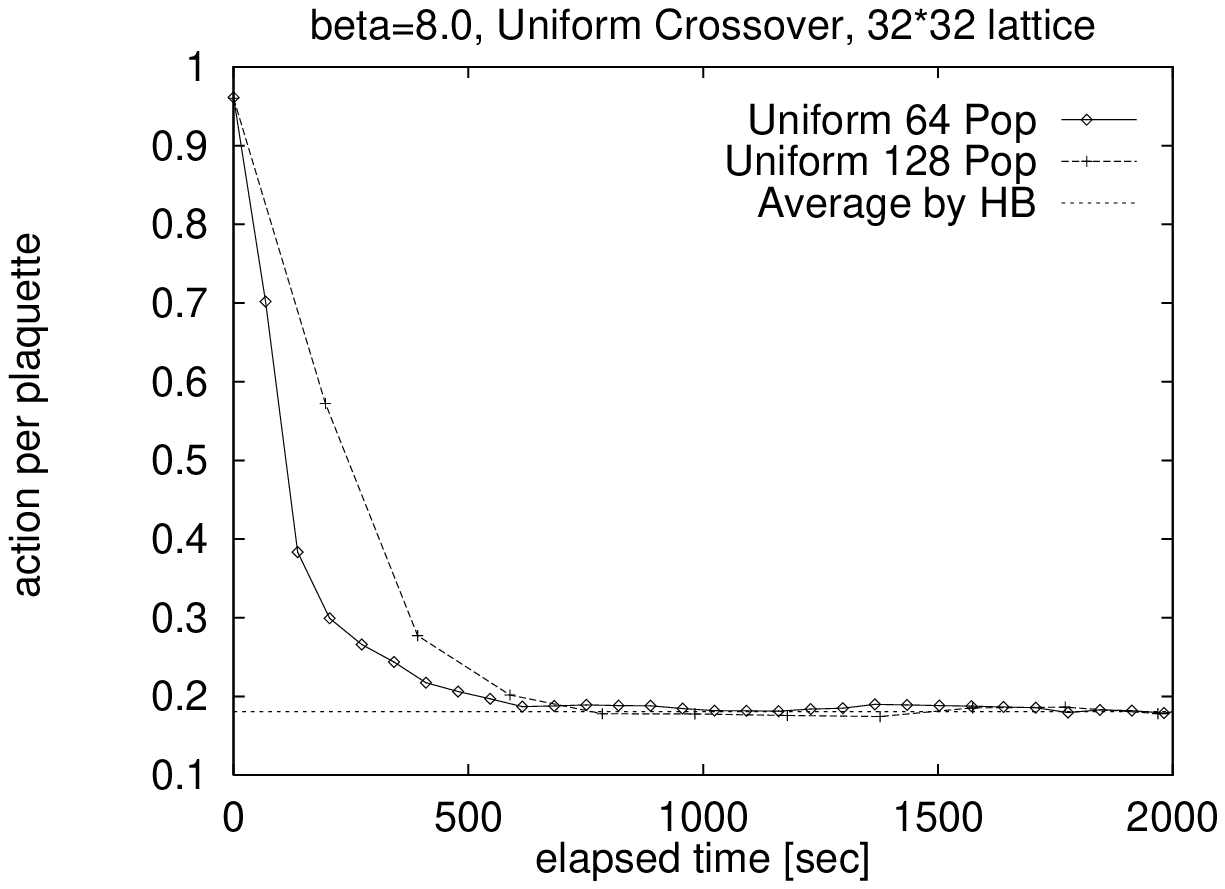,
        width=6.5cm,height=7.0cm,
        bbllx=40pt,bblly=40pt,bburx=340pt,bbury=340pt,
        angle=0}
\end{center}
\end{minipage}&
\begin{minipage}{7.5cm}
\begin{center}
\epsfig{file=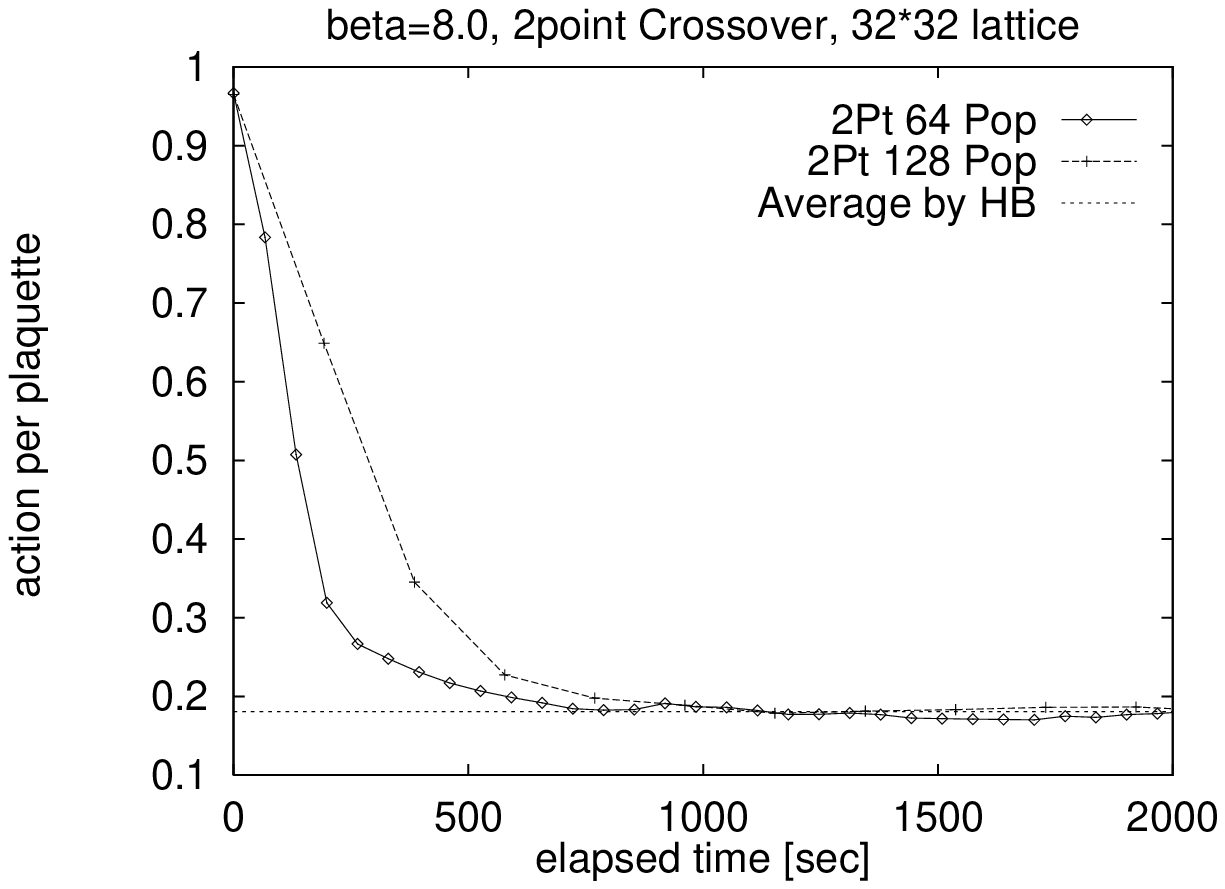,
        width=6.5cm,height=7.0cm,
        bbllx=40pt,bblly=40pt,bburx=340pt,bbury=340pt,
        angle=0}
\end{center}
\end{minipage} \\
\begin{minipage}{7.5cm}
\begin{center}
\epsfig{file=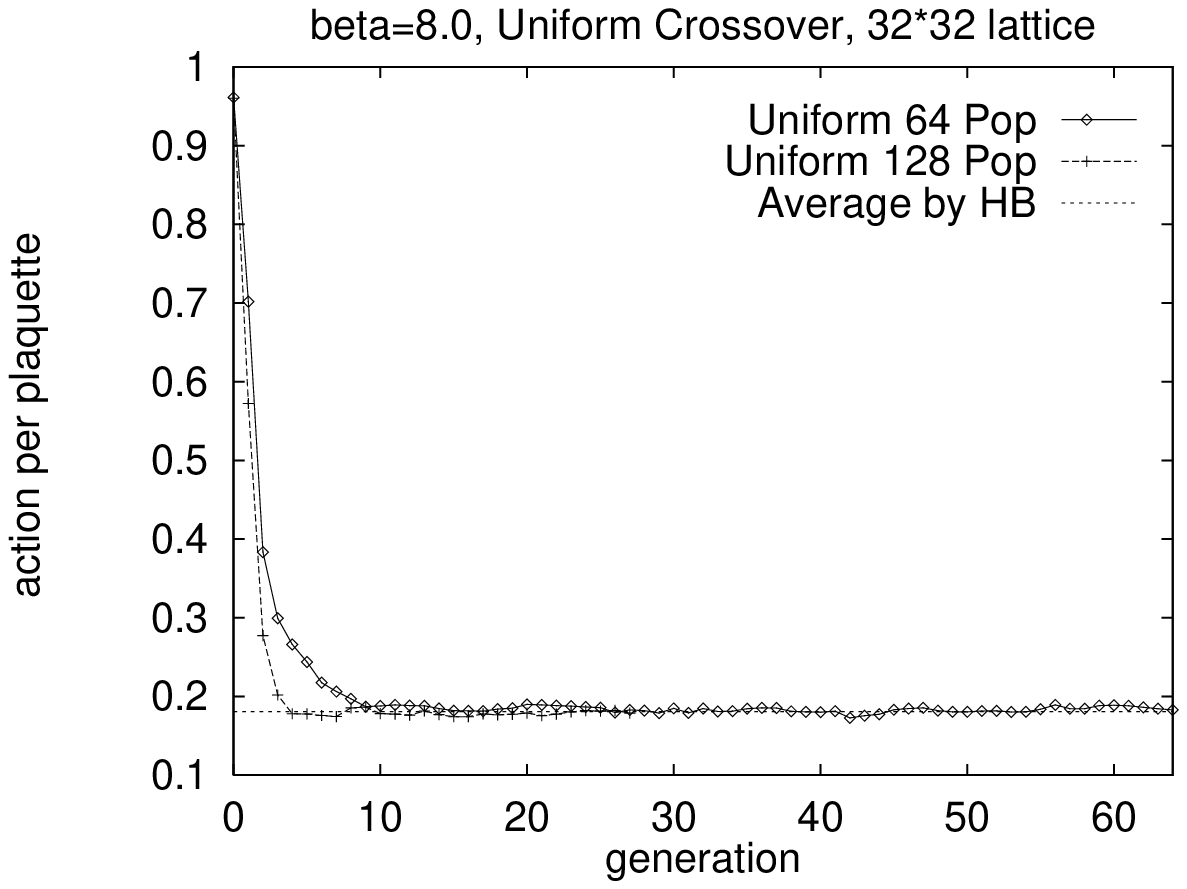,
        width=6.5cm,height=7.0cm,
        bbllx=40pt,bblly=40pt,bburx=340pt,bbury=340pt,
        angle=0}
\end{center}
\end{minipage}&
\begin{minipage}{7.5cm}
\begin{center}
\epsfig{file=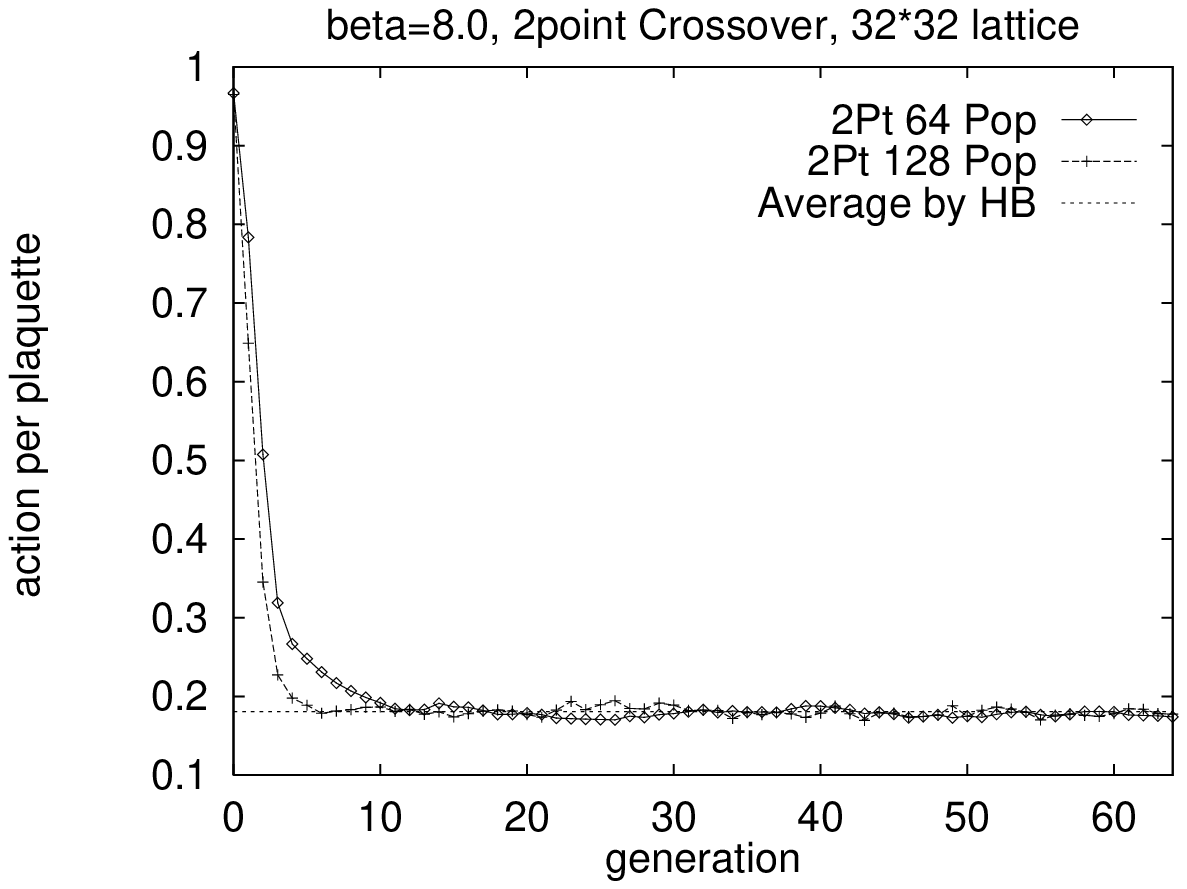,
        width=6.5cm,height=7.0cm,
        bbllx=40pt,bblly=40pt,bburx=340pt,bbury=340pt,
        angle=0}
\end{center}
\end{minipage}
\end{tabular}
\parbox{16cm}{ \caption{ \label{fig5}
Comparison of Recombination schemes and population sizes at
$\beta=8.0$}}
\end{center}
\end{figure}

Figures~\ref{fig4} and ~\ref{fig5} 
show the comparison of schemes and population sizes 
with $\beta$ values $2.0$ and $8.0$, respectively.
In both crossover schemes, there is no difference of convergence speed between
population sizes on the point of elapsed time. On the point of the number of
generation steps, however, the cases of large population size obviously brings
quick decrease of the action  rather than them of small population size.
This is reasonable on GA simulation, since the richer diversity
avoids the capture into a local minima, leading a global minima immediately.
The slow speed of a large population size comes 
from the large number of procedures, because we have not yet 
optimized GAs procedures enough. Simulation with uniform crossover 
scheme does not converge fast.
It is because this scheme brings the increase of the action
caused by the increase of the surface energy that tends to overcome
the decrease of the interior energy.
The fact suggest that the more effective crossover scheme is needed
for the simulation on the higher dimensions.

\section{Conclusion}

\begin{figure}
\begin{center}
\epsfig{file=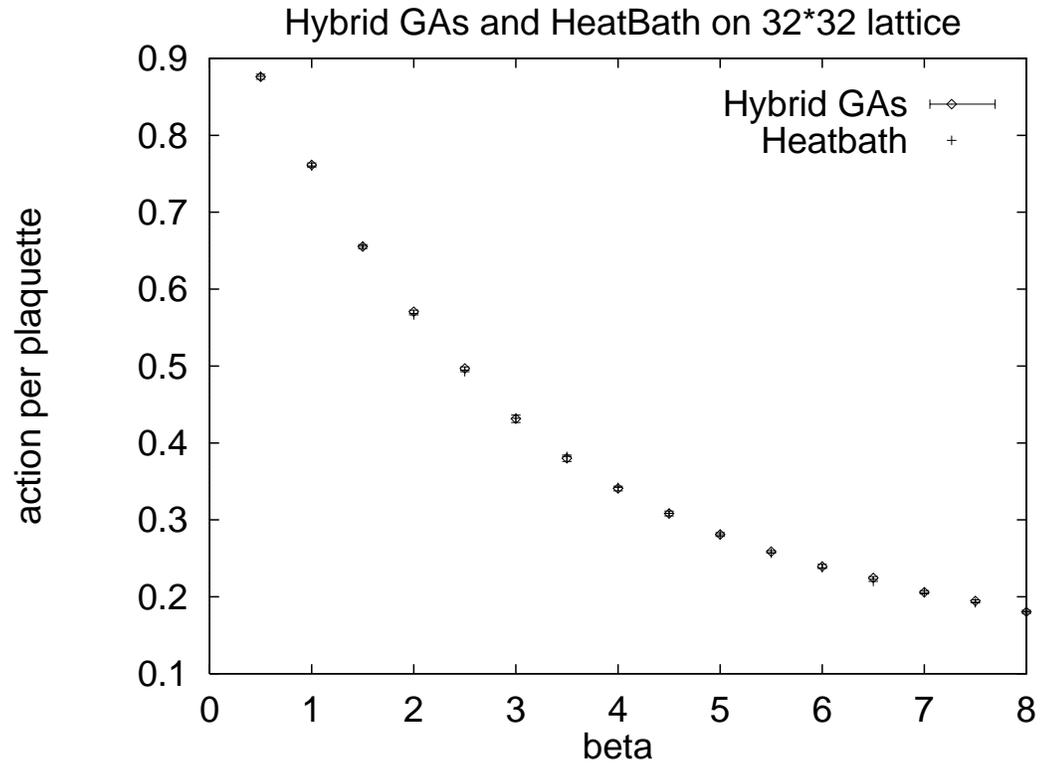,
        width=12.0cm,height=12.0cm,
        bbllx=50pt,bblly=50pt,bburx=350pt,bbury=350pt,
        angle=0}
\end{center}
\begin{center}
\parbox{12.5cm}{ \caption{ \label{fig6}
Action per plaquette given by Hybrid GAs and HeatBath
.}}
\end{center}
\end{figure}

We apply GAs to SU(N) lattice gauge theory. 
The GAs is an extended version of Metropolis method.
We present in this paper an experimental result of SU(2) pure gauge theory on 
a 2 dimensional lattice. Fig ~\ref{fig6} shows the average values 
obtained by GAs and those given by HB for the sake of comparison. 
Schemes, crossrate and mutation rate are same as in the previous section, and
the 2point crossover scheme is used.
Runs for small $\beta$s ($\le 2.0$) have 32 population size and end at 64th
generation. Among them, last 32 generation values are averaged ones.
Simulations for $\beta$s ($\ge 2.5$) have 128 population size 
and end 32th generation. Last 16 generation values are averaged ones.

GAs' simulation converges rapidly at the minimum value of HB,
especially for large $\beta$s.
As shown before,
 for small $\beta$s, fluctuations are, however,very hard at
 every generation step.
If taking into account  that one generation step corresponds
to $128$ sweeps of MPs, their fluctuations become milder and acceptable.

We optimized memory size for the simulation with a large
population size on a work station.
For example, our simulation with 128 population size of $32\cdot32$
lattice needs a memory size not 128 times but only 8 times as in the 
usual methods with the same lattice size.

GAs are effective methods particularly for parallel processing.
Two ways of parallelization are possible, one is a GAs scheme
called migration in that individuals growing up on a island migrate to
other island, the other is that a root processor on which selection and
updating are carried, distributes genome to child processors. Both
methods are possible because GAs treat all data as global.

We show the possibility and effectiveness of GAs 
for SU(N) lattice gauge theory without any optimization of schemes.
The more detailed discussion and study about the detailed 
balance should be needed.
Besides them, physical values of Wilson loop and correlation lengths 
will be calculated in the near future.

\section{Acknowledgment}
I acknowledge the use of the Reproduction Plan Language, RPL2 produced
by Quadrastone Limited, in the production of this work.
I acknowledge the use of workstations of Yukawa Institute.
I thank Prof.A.Sugamoto for discussions, for reading the manuscript and for
his help, Prof.H.Nakajima for useful and helpful discussion, Dr.Kamoshita for
his comments, and Dr.M.Feurstein for making me notice the analogy between
GAs and Metropolis.


\end{document}